\documentclass[prl,reprint,superscriptaddress,showpacs,preprintnumbers,nobibnotes,amsmath,amssymb,aps,longbibliography]{revtex4-1}

\usepackage{cancel}
\usepackage{accents}
\usepackage{mciteplus,slashed}
\usepackage{amssymb,cancel,amsmath}
\usepackage{dcolumn}
\usepackage{bm}
\usepackage[caption=false]{subfig}
\usepackage{appendix}
\usepackage{feynmp-auto}
\usepackage[T1]{fontenc}	
\usepackage{csvsimple}
\usepackage[colorlinks=true,citecolor=blue,linkcolor=blue]{hyperref}
\usepackage[section]{placeins}
\usepackage[capitalise]{cleveref}
\usepackage{booktabs}
\usepackage{graphicx}
\usepackage{subfig}
\usepackage{mathrsfs}
\usepackage{multirow}
\usepackage[utf8]{inputenc}
\usepackage{siunitx}
\usepackage{comment}
\usepackage{gensymb}
\usepackage{enumitem}

\usepackage[dvipsnames]{xcolor}
\usepackage[normalem]{ulem}

\begin{document}

\title{The Neutrino Kaleidoscope: Searches for Non-Standard Neutrino Oscillations at Neutrino Telescopes with a TeV Muon Accelerator Source}

\author{Nicholas W. Kamp}
\affiliation{Department of Physics and Laboratory for Particle Physics \& Cosmology, Harvard University}

\author{Gray Putnam}
\affiliation{Fermi National Accelerator Laboratory, Batavia, Illinois 60510, USA}

\begin{abstract}
Muon accelerators, a potential technology for enabling $\mathcal{O}($\SI{10}{TeV}$)$ parton center of mass energy collisions, would also source an intense, collimated beam of neutrinos at TeV energies. The energy and size of this beam would be excellently matched as a source for existing and planned neutrino telescopes: gigaton-sized detectors of astrophysical neutrinos at and above TeV energies. In this paper, we introduce the technical considerations and scientific reach of pairing a muon accelerator source of neutrinos with a neutrino telescope detector, a combination we dub the ``Neutrino Kaleidoscope''. In particular, such a pairing would enable searches for non-standard oscillations of the beam neutrinos as they traverse the earth between source and detector. 
These non-standard neutrino oscillations could be sourced by Lorentz invariance violation, which a neutrino kaleidoscope could probe up to the quantum gravity-motivated Planck scale. Such a search would also have a reach on sterile neutrinos orders of magnitude beyond existing terrestrial limits.  Finally, we touch on some of the non-oscillation potential of a neutrino kaleidoscope. 
\end{abstract}

\maketitle


\section{Introduction}

Neutrino telescopes are the largest particle physics detectors ever built.
The IceCube observatory specifically instruments approximately one cubic kilometer of South Pole ice with photo-multiplier tubes, corresponding to a metric gigaton of active mass~\cite{IceCube1}.
These telescopes are designed to observe atmospheric and astrophysical neutrino interactions with energies from $\mathcal{O}(10\,{\rm GeV})$~\cite{IceCube:2011ucd} up to $\mathcal{O}(1\,{\rm EeV})$~\cite{IceCube:2025ezc,KM3NeT:2025npi}.
Atmospheric neutrinos with TeV energies have been used by these detectors to search for eV-scale sterile neutrinos~\cite{IceCubeSterileNu1, IceCubeSterileNu2}, Lorentz violation~\cite{AtmospericNeutrinoLIV1, AtmospericNeutrinoLIV2}, and non-standard neutrino-matter interactions~\cite{AtmospericNeutrinoNSI}. A TeV neutrino beam would have the potential to make these searches orders of magnitude more sensitive through higher interaction rates and greater systematic control.

Such a neutrino beam would be sourced by TeV muon accelerators, a novel technology proposed to enable high energy particle collisions with a parton center-of-mass energy $\gtrsim$2-\SI{10}{TeV}~\cite{MuonCollider1, MuonCollider2, MuonCollider3, MuonCollider4, MuonCollider5}. The decay of the beam muons would source an intense, collimated flux of electron and muon (anti-)neutrinos ~\cite{NuPhysNuFact, uCDesign2}. This neutrino flux could be utilized in a variety of measurements by a detector at the accelerator site~\cite{NuPhysMuCollider1, NuPhysMuCollider2, deGouvea:2025zfq}, and also would remain intense even far away from the accelerator~\cite{NeutrinoRadiation, NeutrinoCommunication, NeutrinoBomb}.

\begin{figure}[t]
    \centering
    \includegraphics[width=0.49\textwidth]{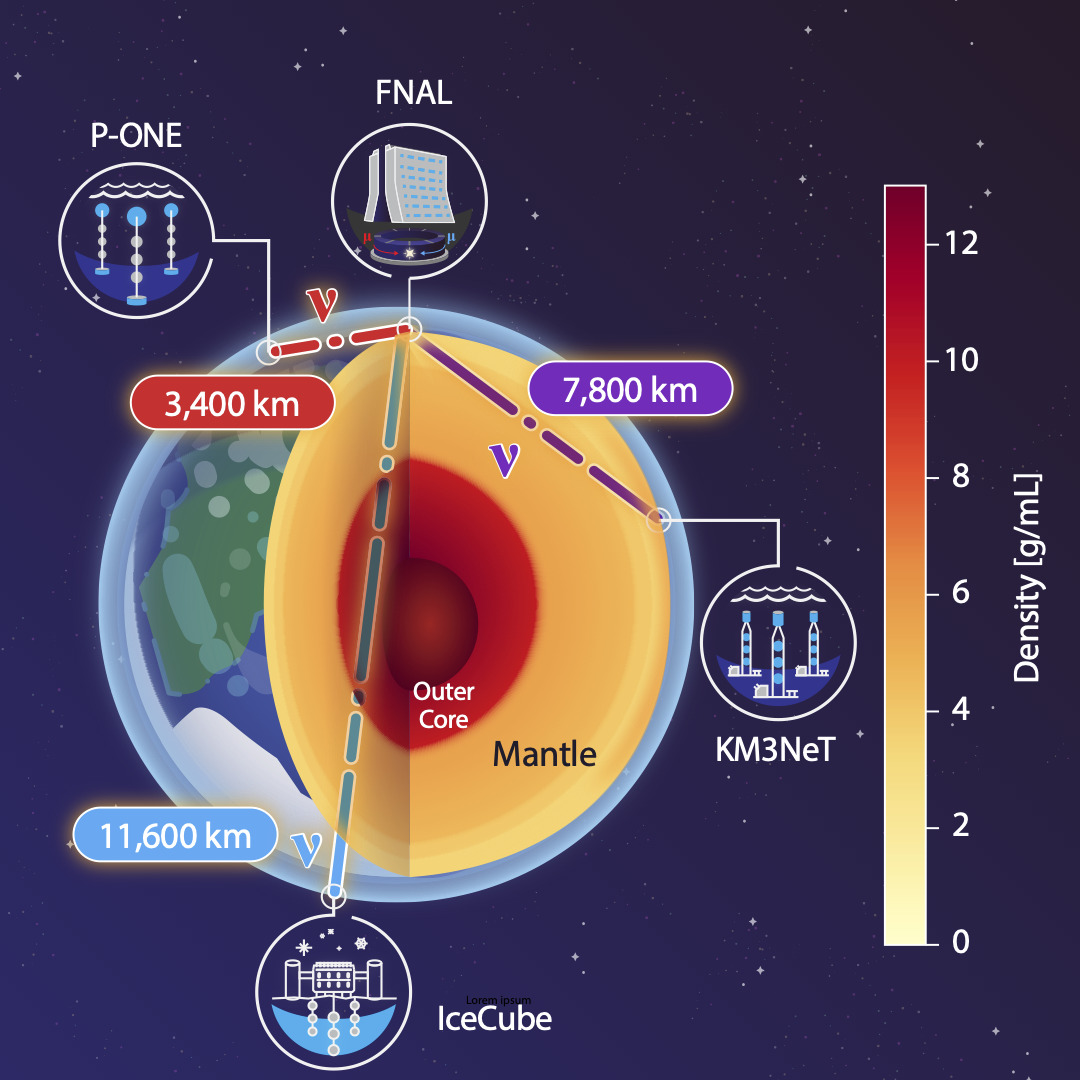}
    \caption{Baseline scenarios of pairings of muon accelerators and neutrino telescopes as a neutrino kaleidoscope. As is discussed in the text, other scenarios have analogous baselines to those depicted here. Density is taken from PREM~\cite{PREM}.}
    \label{fig:kaleidoscope}
\end{figure}

In this letter, we investigate the feasibility and scientific reach of pairing a TeV muon accelerator source of neutrinos with a neutrino telescope detector (see Fig.~\ref{fig:kaleidoscope}). 
We dub this combination the neutrino kaleidoscope (alternatively, ``collide''-o-scope). As long as the beam dispersion is kept small enough, the angular size of the neutrino beam spot is proportional to $1/\gamma$, where $\gamma$ is the boost factor of the muons in decay~\cite{NuPhysMuCollider1}. This angular size translates to a spatial size of $\mathcal{O}$(\SI{100}{\meter}) at the telescope,
well matched to the size of the detector. Furthermore, at these energies the shadowing of the neutrino flux by the earth is small~\cite{NeutrinoEarthShadow, NeutrinoXSecs}, while both the cross section and detector are large enough that a significant fraction of the neutrino flux would interact in the telescope: about 1 in $10^6$ neutrinos, for a \SI{1}{\giga\tonne} detector.

We consider three baseline pairing scenarios of source and detector: Fermilab to P-ONE~\cite{P-ONE}, KM3NeT~\cite{KM3Net}, or IceCube~\cite{IceCube1}. These three scenarios match to three qualitatively different neutrino trajectories through the crust, mantle or core (respectively). They also match well onto other possible pairings with existing~\cite{IceCube1,KM3Net,Baikal-GVD} or proposed~\cite{P-ONE,IceCube-Gen2:2020qha,TRIDENT:2022hql,Zhang:2024slv,Huang:2023mzt} neutrino telescopes, as well as CERN to KM3NeT (crust-crossing) or IceCube (core-crossing).

We focus on the non-standard neutrino oscillation scenarios that could be probed by a neutrino kaleidoscope. At the relevant baseline and energy ($\sim$ \SI{10000}{\kilo\meter} / \SI{1}{TeV}), there are various new physics scenarios that would induce novel neutrino oscillations between the accelerator source and neutrino detector. These scenarios arise from models motivated by the existence of neutrino mass (such as right-handed neutrino states~\cite{AtmoSterileNu1, AtmoSterileNu2} and neutrino mass generation models~\cite{AtmoNSI}), as well as Lorentz invariance violation~\cite{LIVNuOsc, LIVNuOsc1, LIVNuOsc2}, which a neutrino kaleidoscope could probe up to the quantum gravity-motivated Planck scale.

\section{Experimental Concept}
\subsection{Accelerator}
There are two primary requirements that a muon accelerator must meet for it to support a neutrino kaleidoscope. First, some section of the accelerator must be engineered to point at the telescope. This would require achieving a descent angle of $13^\circ$ (Fermilab to P-ONE), $34^\circ$ (Fermilab to KM3NeT), or $66^\circ$ (Fermilab to IceCube).
If a section of the accelerator or collider ring cannot be constructed to point at the destination telescope, one could instead consider pointing the trajectory of the beam as it is dumped.

The second requirement is that the beam divergence of the muon beam must be kept small. 
For a \SI{5}{TeV} accelerator, the inherent spread of the neutrino beam ($1/\gamma$) is about \SI{0.02}{\milli\radian}. Any beam divergence larger than this will attenuate the flux observed in the telescope. Furthermore, there are spectral features in the decay angle that enhance the ability of a neutrino kaleidoscope to search for oscillations (these are similar in nature to the PRISM effect in neutrino super beams~\cite{PRISM}). Muon collider designs have established a goal for the transverse beam emmittance of \SI{25}{\micro\meter\radian}~\cite{uCDesign1, uCDesign2}. For a beam spot size of \SI{5}{\milli\meter} (away from the interaction point), this corresponds to a beam divergence of $5/\gamma\,$\si{\milli\radian}~\cite{TransverseBeam}, or \SI{0.1}{\micro\radian} for a \SI{5}{TeV} muon beam. In the vicinity of the interaction point, the beam is squeezed into a smaller spot size which corresponds to a much larger divergence and therefore a less useful neutrino beam~\cite{NuPhysMuCollider1, uCDesign2}. Outside the interaction point, movers will likely be necessary to spread the neutrino beam by about \SI{1}{\milli\radian} to reduce the impact of neutrino radiation near the accelerator~\cite{uCDesign2}. Although this will reduce the total neutrino flux, the instantaneous (turn-by-turn, i.e.) beam divergence would still be \SI{0.1}{\micro\radian}, and the key spectral features in the beam would be retained.

We estimate a baseline number of muon decays from a few scenarios for the beam. In the first, the beamline is engineered so that a single straight section is pointed to the telescope. For a baseline muon accelerator scenario~\cite{uCDesign1,uCDesign2} -- $2\times10^{12}$ muons per bunch, a \SI{10}{\kilo\meter} circumference ring, \SI{30}{\centi\meter} straight section, spacer reduction of $50\times$ -- there would be $5\times 10^{14}$ muon decays per year of total beam operation. In the second, the beam dump is directed towards the telescope. For a \SI{5}{\hertz} repetition (and dump) rate and a \SI{25}{\meter} decay tunnel (that is not subject to movers), there would be $2\times 10^{14}$ muon decays per year of total beam operation. We consider the smaller beam dump scenario ($2\times 10^{14}$) as a nominal yearly rate.
As an alternative scenario, the interaction point (IP) could be oriented towards the detector. This would enable a larger \SI{100}{\meter} straight section with a intrinsic angular spread of \SI{0.6}{\milli\radian} (due to the spatial focusing at the IP~\cite{uCDesign2}). In this case, there would be $8\times10^{18}$ $\mu$ decays per year, reduced by $30\times$ by the angular spread, and with the spectral features lost. We consider this scenario for only the P-ONE (crust-crossing) scenario, due to the challenge of orienting the slope of the IP. 

\subsection{Telescope}

Neutrino telescopes are an ideal far detector for the intense flux of TeV energy neutrinos from a muon collider~\footnote{The ORCA detector of the KM3NeT experiment has previously been considered as a far detector option in a low-energy neutrino beamline~\cite{Singha:2022btw}. Our proposal is quite different in that it benefits from the higher intensities, energies, and collimation of the neutrino flux from a muon collider.}.
The extreme collimation of these neutrinos means that the majority of the flux is contained within $\mathcal{O}(100\,{\rm m})$ over Earth-diameter-scale baselines; thus, we benefit directly from the large active mass of neutrino telescopes.
Neutrino interactions in these detectors generally fall under two morphological categories: \textit{tracks} from high-energy muons and and \textit{cascades} from hadronic and electromagnetic showers.
For a $\mu^+$ ($\mu^-$) accelerator, charged-current $\bar{\nu}_\mu$ ($\nu_\mu$) interactions would result tracks, while charged-current $\nu_e$ ($\bar{\nu}_e$) interactions and neutral-current interactions of all flavors would result in cascades.
Track samples tend to have a $\overset{(-)}{\nu}_\mu$ purity $\gtrsim 99\%$~\cite{IceCubeCollaboration:2024dxk}, while cascade samples are comprised of similar contributions from all flavors~\cite{IceCube:2023ame}.
Tracks have great angular resolution ($\lesssim 1\degree$~\cite{AMANDA:2003vtt}) but relatively poor resolution on the muon energy ($\sim 0.22$ in $\log_{10}(E_\mu)$~\cite{IceCube:2012iea}).
Cascades have much better resolution on the deposited energy ($\sim 0.1$ in $\log_{10}(E)$~\cite{Abbasi:2021ryj}).
This can also improve the energy reconstruction of $\nu_\mu$ interactions inside the detector, which create a hadronic cascade at the start of the track~\cite{IceCubeCollaboration:2024dxk}.

\section{Sterile Neutrino Oscillations}

A new right-handed (non-weakly interacting, or sterile) neutrino state  would induce oscillations from the active beam neutrinos into the sterile state. The energy and baseline of neutrino kaleidoscopes would be precisely matched to probe oscillations at an active-sterile mass splitting of about \SI{1}{eV\squared}. In particular, at TeV neutrino energies and eV$^2$ splittings there is a matter resonance, most prominent through the core of the earth, that would magnify the oscillation of muon anti-neutrinos and electron neutrinos into the sterile state~\cite{AtmoSterileNu1, AtmoSterileNu2, MSW1, MSW2, MSW3}. 


An active-sterile oscillation can be described in the ``3+1'' picture 
as
\begin{equation}
\begin{split}
    &i\frac{d}{dx}\begin{pmatrix}
        \nu_\alpha\\
        \nu_S
    \end{pmatrix} = \\
    &\quad\begin{pmatrix}
        -\frac{\Delta m^2}{4E}\cos 2\theta_{\alpha\alpha} \pm_\alpha G_F N_e/\sqrt{2} & \frac{\Delta m^2}{4E}\sin 2\theta_{\alpha\alpha} \\ 
        \frac{\Delta m^2}{4E}\sin 2\theta_{\alpha\alpha} & \frac{\Delta m^2}{4E}\cos 2\theta_{\alpha\alpha}
    \end{pmatrix}\begin{pmatrix}
        \nu_\alpha\\
        \nu_S
    \end{pmatrix}
        ,\,        
\end{split}
\label{eq:sterilenu}
\end{equation}
where $\alpha \in \{e, \mu\}$ is either an electron or muon neutrino, $S$ represents the sterile state, $\theta_{\alpha\alpha}$ is the effective active-sterile neutrino mixing angle (in terms of the mass-mixing matrix $U$, $\sin^22\theta_{\alpha\alpha}\equiv 4|U_{\alpha}|^2 [1 - |U_{\alpha}|^2]$), $\Delta m^2$ is the new mass splitting, $G_F$ is the Fermi constant, and $N_e(\approx N_n)$ is the number density of electrons~\cite{SterileNuMatter}. The sign of the matter potential term $(\pm_\alpha)$ is $+1$ for electron neutrinos and muon anti-neutrinos, and $-1$ for electron anti-neutrinos and muon neutrinos. When the matter potential is positive, the resonance is achieved when $\frac{\Delta m^2}{4E}\cos 2\theta_{\alpha\alpha} = G_F N_e/\sqrt{2}$. 

\begin{figure}[t]
    \centering
    \includegraphics[width=0.49\textwidth]{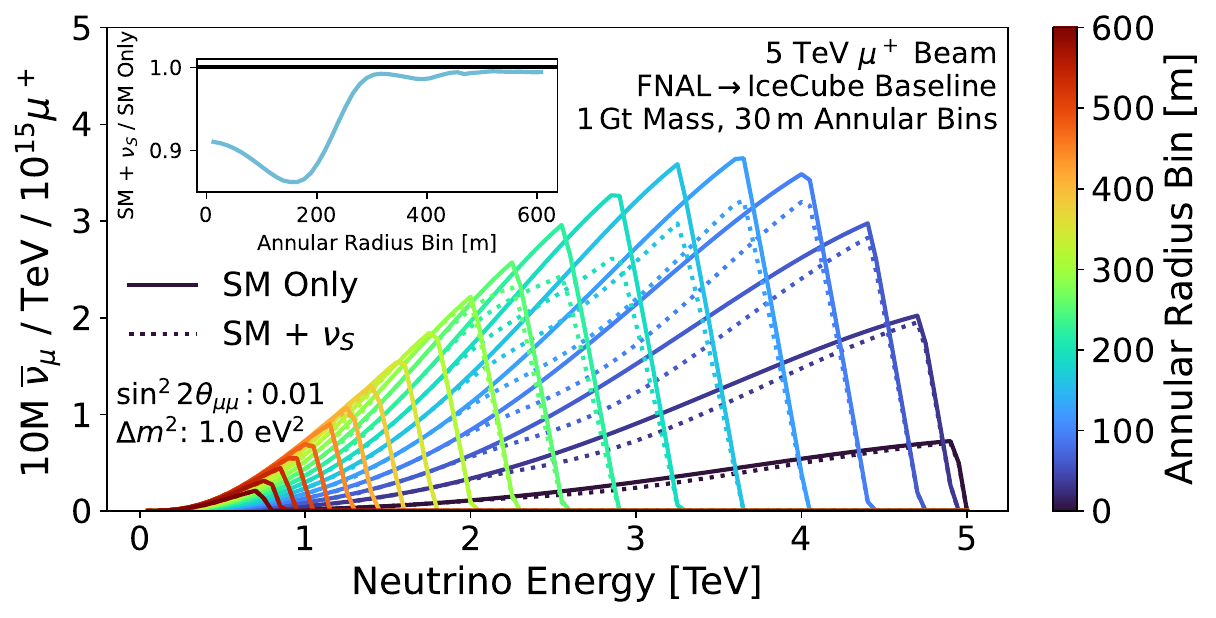}
    \caption{Neutrino interaction rates, per \SI{30}{\meter} annular radius bin, at the Fermilab to IceCube baseline. Shown with and without an example injected sterile neutrino ($\nu_S$) model point. The inset shows the fractional rate of $\overline{\nu}_\mu$ interactions with the sterile neutrino (relative to without) as a function of the annular radius.}
    \label{fig:sterilenu_distributions}
\end{figure}

Fig.~\ref{fig:sterilenu_distributions} shows the distribution of muon anti-neutrino events by the neutrino energy and interaction radius at the IceCube baseline, with an example sterile neutrino point injected. Because different annular radii contain different energy distributions, an oscillation effect is observable even in the vertex position of the neutrino interaction. This is useful because reconstructing the energy of $\mathcal{O}$(TeV) ``track-like'' neutrino interactions is challenging in neutrino telescopes~\cite{IceCube:2012iea}. 

\begin{figure}[t]
    \centering
    \includegraphics[width=0.49\textwidth]{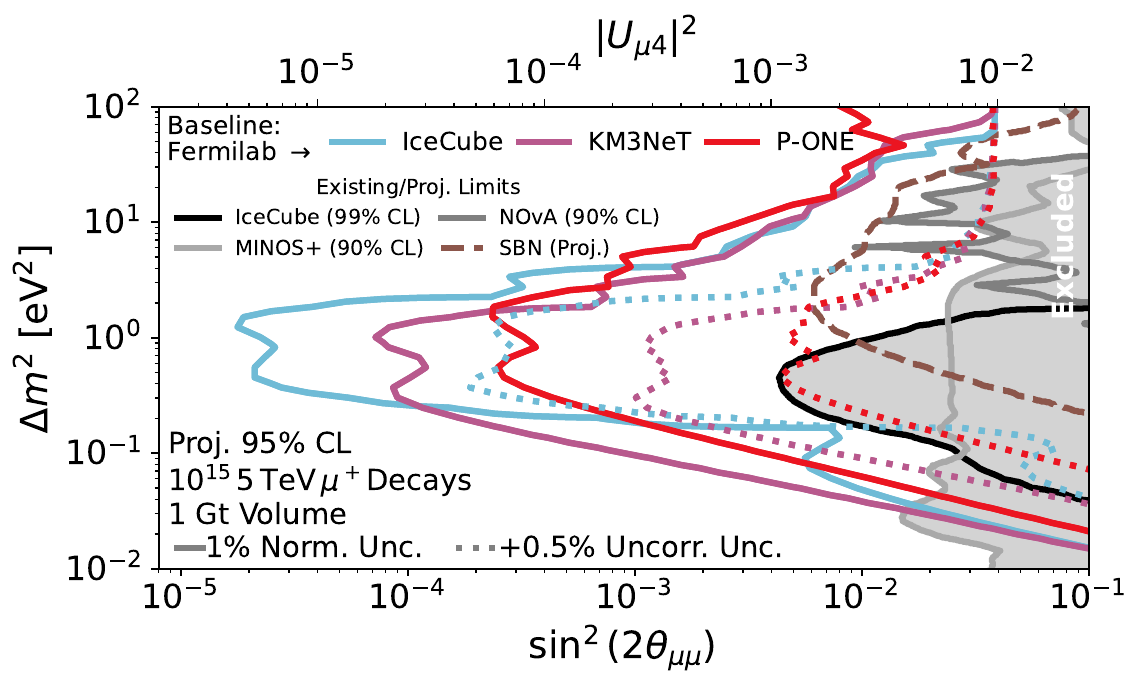}
    \caption{Projected sensitivity of a neutrino kaleidoscope to an eV$^2$-scale, muon-coupled sterile neutrino, compared to existing and projected terrestrial limits~\cite{MINOSSterileNu, SBNSterileNu, IceCubeSterileNu2, NOvASterileNu}. Shown at various source-detector baselines with an integrated flux of $10^{15}$ $\mu^+$ decays at an energy of \SI{5}{TeV}. The sensitivity is calculated with a 1\% normalization uncertainty (solid line) and an additional 0.5\% uncorrelated systematic uncertainty in each annular bin (dashed line). }
    \label{fig:sterilenu_sensitivity}
\end{figure}

We estimated the sensitivity of neutrino kaleidoscopes with with a binned likelihood analysis performed on the neutrino interaction annular radius in bins of \SI{10}{\meter}~\cite{vanEeden:2021onr}.
Oscillations are computed by numerically integrating Eq.~\ref{eq:sterilenu} with the earth density taken from PREM~\cite{PREM}. The resulting sensitivity is shown in Fig.~\ref{fig:sterilenu_sensitivity}. 
Depending on the baseline and systematic uncertainties, the sensitivity reaches up to multiple orders of magnitude beyond any existing or planned search for \si{eV}-scale sterile neutrinos. 

The incredible reach of neutrino kaleidoscopes to sterile neutrinos is underlaid by numerous advantages over existing searches. The neutrino energy is at the \si{TeV} scale, and is therefore not subject to the large systematic uncertainties that impact \si{GeV}-scale neutrino experiments~\cite{NuSTEC}. The search also gets a significant boost from the matter resonance which, unlike in searches with \si{TeV} atmospheric neutrinos, is probed with a pure $\overline{\nu}_\mu$ beam with all of the neutrinos experiencing the same density profile. Finally, the rates in the the telescope are large enough (100s of millions) to search for very small oscillation effects, as long as systematic effects can be mitigated.

Neutrino kaleidoscopes would also have a similar reach to electron-coupled sterile neutrinos, which would also probe orders of magnitude beyond existing searches~\cite{PROSPECTSterileNu, Prospect2, STEREOSterileNu, DayaBaySterileNu, BESTSterileNu}. In this case, the ``shower-like'' reconstructed topology would include both $\nu_e$ charged-current and neutral-current events. The energy of the neutrino could also be reconstructed with better precision, complementing the radius reconstruction.

\section{Active Neutrino Oscillations from Quantum Gravity} \label{sec:lv}

Neutrino kaleidoscopes would also be able to probe the oscillation between the active neutrino states in the beam. At \si{TeV} energies, the matter potential in the earth effectively decouples the $\nu_e$ flavor from $\nu_\mu$ and $\nu_\tau$. Including the possibility of a non-standard neutrino potential $V$, the $\mu-\tau$ oscillation master equation is  
\begin{equation}
\begin{split}
    &i\frac{d}{dx}\begin{pmatrix}
        \nu_\mu\\
        \nu_\tau
    \end{pmatrix} = \\
    &\quad\left[\frac{\Delta m_{31}^2}{4E}\begin{pmatrix}
        -\cos 2\theta_{23} & \sin 2\theta_{23} \\ 
        \sin 2\theta_{23} & \cos 2\theta_{23}
    \end{pmatrix}+\begin{pmatrix}
        V_{\mu\mu} & V_{\mu\tau} \\ 
        V^*_{\mu\tau} & V_{\tau\tau}
    \end{pmatrix}\right]\begin{pmatrix}
        \nu_\mu\\
        \nu_\tau
    \end{pmatrix}
        ,\,        
\end{split}
\end{equation}
where $\theta_{23}$ and $\Delta m_{31}^2$ are the standard neutrino oscillation parameters. The sign of the potential $V$ is flipped for anti-neutrinos. In the limit where the baseline is short ($\frac{\Delta m_{31}^2}{4E}L, VL << 1$), for a position-independent potential, the transition probability becomes
\begin{equation} \label{eq:active_osc}
    P_{\nu_\mu\to\nu_\tau} = L^2\left|\frac{\Delta m_{31}^2}{4E}\sin 2\theta_{23} + V_{\mu\tau}\right|^2 + \mathcal{O}(L^4).\,
\end{equation}
There are therefore two predominant channels to observe active neutrino oscillations: muon neutrino disappearance and tau neutrino appearance. 

The potential $V_{\mu\tau}$ in \cref{eq:active_osc} can be generated by theories of Quantum Gravity (QG)~\cite{Hawking:1982dj}.
QG theories unify general relativity with the Standard Model at the Planck scale ($E_P = 1.22 \times 10^{19}\,{\rm GeV}$), roughly 17 orders of magnitude above the electroweak scale~\cite{Kostelecky:1988zi}.
While Planck-scale QG cannot presently be probed directly, it can lead to Planck-suppressed effects on observables at lower energy scales~\cite{Addazi:2021xuf}, including small violations of Lorentz symmetry~\cite{Colladay:1998fq,Kostelecky:1988zi,Kostelecky:2008ts,Pospelov:2010mp}.
Notably, non-standard neutrino oscillations are a powerful probe of such Planck-suppressed Lorentz violation~\cite{Coleman:1998ti,LIVNuOsc,LIVNuOsc1,Kostelecky:2003cr,Kostelecky:2004hg}.
The strongest constraints on QG-induced Lorentz violation in the neutrino sector come from IceCube measurements of atmospheric~\cite{IceCube:2010fyu,IceCube:2017qyp,Skrzypek:2024vqm} and astrophysical~\cite{IceCube:2021tdn} neutrinos, as they probe the highest energies and longest baselines of all neutrino experiments.
Neutrino kaleidoscopes have typical energies and baselines similar to that of IceCube atmospheric neutrino searches, with the added benefit of unprecedented event rates.
Thus, they are an ideal experimental configuration to search for QG-induced Lorentz violation.

The imprint of Planck-scale Lorentz violation on the Standard Model is described by an effective field theory, the Standard-Model Extension (SME)~\cite{Colladay:1996iz,Colladay:1998fq}, which includes operators that govern neutrino interactions with a new Lorentz-violating background field.
The renormalizable $\mu$-$\tau$ operators of this group are of mass dimension three and four and are parameterized by the Wilson coefficients $(a_L^\mu)_{\mu \tau}$ and $(c_L^{\mu \nu})_{\mu \tau}$, respectively.
Unlike typical Wilson coefficients, these carry Lorentz indices, which leads to frame-dependent neutrino interactions.
This effect is conventionally studied within a Sun-centered celestial-equatorial frame, in which the Earth's rotation causes the direction of a neutrino beam to vary over the course of a sidereal day, i.e. with a frequency $\omega_{\oplus} = 2\pi/(23~{\rm h}~56~{\rm min})$~\cite{Kostelecky:2004hg}.
This leads to a potential $V_{\mu \tau}$ with an explicit sidereal time dependence,
\begin{equation} \label{eq:lv_sidereal}
\begin{split}
V_{\mu \tau} = 
&(\mathcal{C})_{\rm \mu \tau}
 +  (\mathcal{A}_s)_{\rm \mu \tau} \sin \omega_\oplus T_\oplus 
 +  (\mathcal{A}_c)_{\rm \mu \tau} \cos \omega_\oplus T_\oplus \\
 +  &(\mathcal{B}_s)_{\rm \mu \tau} \sin  2 \omega_\oplus T_\oplus 
 +  (\mathcal{B}_c)_{\rm \mu \tau} \cos 2 \omega_\oplus T_\oplus,
 \end{split}
\end{equation}
where, dropping the flavor indices for brevity, $\mathcal{C}$, $\mathcal{A}_s$, $\mathcal{A}_c$, $\mathcal{B}_s$, and $\mathcal{B}_c$ are energy-dependent linear combinations of the $a_L^\mu$ and $c_L^{\mu \nu}$ coefficients.
The specific linear combination depends on the geometry of the neutrino beam with respect to the Earth; explicit expressions can be found in Eq.~4-13 of Ref.~\cite{Kostelecky:2004hg}.

Sidereal variations from Lorentz violation have been perviously investigated in neutrino experiments~\cite{IceCube:2010fyu,MINOS:2008fnv,MINOS:2010kat,LSND:2005oop,MiniBooNE:2011pix}.
In this Letter, we study the expected sensitivity of neutrino kaleidoscopes to sidereal variations.
This analysis benefits directly from the high event rates of neutrino kaleidoscopes, as systematic uncertainties tend to be negligible~\cite{MINOS:2008fnv,IceCube:2010fyu}.
We consider the beam dump scenario for IceCube and KM3NeT and the \SI{100}{\meter} IP straight section scenario for P-ONE, denoted P-ONE (IP), as energy-radial correlations are not important in the sidereal variation analysis.
We focus on $\nu_\mu$ disappearance from $\mu$-$\tau$ SME coefficients, as tracks provide a pure sample of $\nu_\mu$ interactions.
We also consider $\nu_\tau$ appearance in hypothetical sample of $\nu_\tau$ interactions.
While event-by-event $\nu_\tau$ identification is difficult at TeV energies~\cite{IceCube:2020fpi}, techniques such as the neutron/muon echo~\cite{Li:2016kra} and the isolation high-angle muons from tau decay may enable the identification of a $\nu_\tau$-enhanced sample.
As a benchmark, we consider signal-to-background ratio of 0.1 for this $\nu_\tau$-enriched sample.

\begin{figure}
    \centering
    \includegraphics[width=0.5\textwidth]{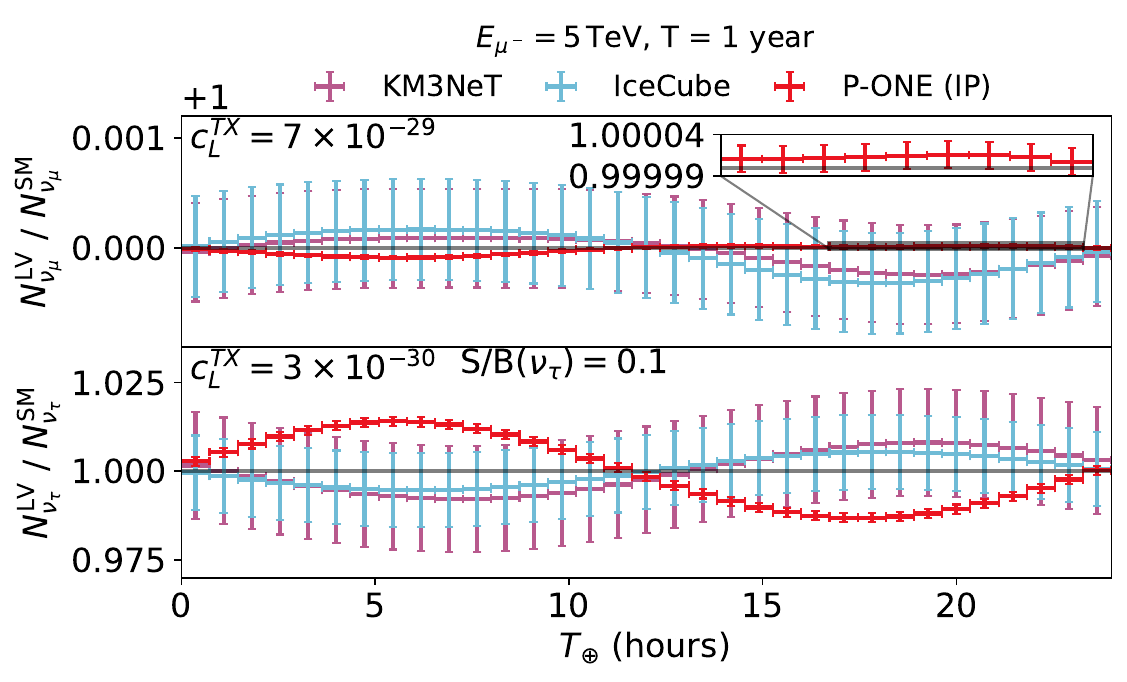}
    \caption{The ratio of muon (top) and tau (bottom) neutrino interactions with and without Lorentz violation as a function of sidereal time, considering a single nonzero SME Wilson coefficient. Error bars indicate the expected statistical uncertainty after one year of accelerator operation.}
    \label{fig:lv_example}
\end{figure}

\Cref{fig:lv_example} shows the expected interaction rate in each detector for the $\nu_\mu$ and $\nu_\tau$-enriched sample over one sidereal day for characteristic values of $c_L^{TY}$, as a ratio to the Standard Model expectation.
We perform the Fourier analysis outlined in Ref.~\cite{MINOS:2008fnv} to asses our sensitivity to the $a_L^\mu$ and $c_L^{\mu \nu}$ coefficients.
\Cref{tab:liv_constraints} suggests that the $\nu_\mu$ ``track'' sample of a neutrino kaleidoscope can improve upon current limits from IceCube~\cite{IceCube:2010fyu} and MINOS~\cite{MINOS:2010kat} by two or more orders of magnitude.

Many theories of QG~\cite{Berger:2001rm,Bolokhov:2005cj,Carroll:2001ws,Pospelov:2010mp} predict higher-order Lorentz-violating operators with mass dimension $d>4$.
The Planck-scale nature of these theories implies that the SME coefficients associated with higher-order operators have a value of approximately $E_P^{4-d}$.
Furthermore, these operators induce oscillations at a frequency proportional to $L E^{d-3}$~\cite{Kostelecky:2011gq}, such that high-energy neutrinos are more sensitive to higher-dimension operators.
Recent IceCube constraints using high-energy astrophysical neutrino flavor have reached the $E_p^{-2}$ regime of the isotropic $d=6$ operators for the first time~\cite{IceCube:2021tdn}.
However, these constraints depend on the unknown flavor composition of neutrinos at the source.
Neutrino kaleidoscopes can probe higher-dimensional Lorentz-violating operators without such model dependence.

We can estimate the sensitivity to higher-dimensional operators by the relation~\cite{Kostelecky:2011gq}
\begin{equation} \label{eq:lv_higher_order}
|\mathbf{p}|^{d-3} |A_c^d| \approx |A_c^3|,
\end{equation}
where $\mathbf{p}$ is the average neutrino energy, $A_c^3$ is the dimension three coefficient appearing in \cref{eq:lv_sidereal}, and $A_c^d$ is an experiment-dependent linear combination of $d$-dimensional SME coefficients.
We consider three different experimental configurations to probe dimension six operators,
\begin{itemize}
    \item \text{Nominal:} $\nu_\mu$ sample, $T=1\,{\rm yr}$, $E_\mu=5\,{\rm TeV}$
    \item \text{Tau-enriched:} $\nu_\tau$-enriched sample with a signal-to-background ratio of 10\%, $T=10\,{\rm yrs}$, $E_\mu=5\,{\rm TeV}$
    \item \text{High-energy:} $\nu_\mu$ sample, $T=10\,{\rm yrs}$, $E_\mu=50\,{\rm TeV}$
\end{itemize}
\Cref{tab:lv_higher_order} shows that the tau-enriched and high-energy
neutrino kaleidoscope configurations can reach the double-Planck-suppressed regime of $A_c^6$.
Thus, a neutrino kaleidoscope with either a 50 TeV muon beam or improved $\nu_\tau$ tagging at the detector is likely the only terrestrial (and thus model-independent) probe of the double-Planck suppressed imprint of Quantum Gravity.

\begin{table}[]
    \centering
    \begin{tabular}{|c|c|c|c|c|}
Coefficient & KM3NeT & IceCube & P-ONE (IP) & Current Limit \\
 \hline
 $a_L^X / 10^{-25}\,{\rm GeV}$ & 8.2 & 5.8 & 1.0 & $1.7 \times 10^{2}$ (\cite{IceCube:2010fyu}) \\
$c_L^{TX} / 10^{-28}$ & 1.6 & 1.1 & 0.2 & $79$ (\cite{IceCube:2010fyu}) \\
$c_L^{XX} / 10^{-28}$ & 4.0 & 2.4 & 0.7 & $2.5 \times 10^{5}$ (\cite{MINOS:2010kat}) \\
    \end{tabular}
    \caption{The sensitivity of the $\nu_\mu$ sample in each kaleidoscope configuration to three different $a_L$ and $c_L$ SME coefficients at the $3\sigma$ C.L., considering one year of data at a muon energy of $5\,{\rm TeV}$. 
    The current limits on each coefficient are also shown, which come from sidereal variation searches at either IceCube~\cite{IceCube:2010fyu} or MINOS~\cite{MINOS:2010kat}.}
    \label{tab:liv_constraints}
\end{table}

\begin{table}[]
    \centering
    \begin{tabular}{|c|c|c|c|}
Configuration & KM3NeT & IceCube & P-ONE (IP) \\
\hline
Nominal & $2.8 \times 10^{3} / E_P^{2}$ & $1.2 \times 10^{3} / E_P^{2}$ & $5.3 \times 10^{2} / E_P^{2}$ \\
Tau-enriched & $26 / E_P^{2}$ & $17 / E_P^{2}$ & $2.1 / E_P^{2}$ \\
High-energy & $11 / E_P^{2}$ & $5.3 / E_P^{2}$ & $1.9 / E_P^{2}$ \\
    \end{tabular}
    \caption{The sensitivity of each kaleidoscope to $\mathcal{A}_c^6$, in units of $E_P^{-2} = 6.7 \times 10^{-39}\,{\rm GeV}^{-2}$, at the 90\% C.L., considering the three different experimental configurations in the text.} 
    \label{tab:lv_higher_order}
\end{table}

\section{Conclusion}

We have introduced the “neutrino kaleidoscope,” combining a TeV-scale muon accelerator with a gigaton neutrino telescope to deliver, for the first time, a high-energy, long-baseline neutrino experiment with an intense and precisely controlled flux.
The unprecedented event rates enable orders-of-magnitude improvement over existing and planned experiments in sensitivity to non-standard neutrino oscillations from eV-scale sterile neutrinos and Lorentz-violating effects.
Notably, this approach provides the first terrestrial access to double-Planck suppressed sidereal variations from Lorentz violation in theories of quantum gravity.
The neutrino kaleidoscope thereby offers a powerful, generalizable platform to probe new physics and fundamental symmetries with precision.
We encourage further exploration of this paradigm as a flagship opportunity for future accelerator and neutrino detector facilities.

\section{Acknowledgments}

The authors would like to thank Carlos Argüelles, Cari Cesarotti, Joseph Zennamo and Diktys Stratakis  for helpful discussions in the preparation of this manuscript.
The authors are grateful to Jackapan Pairin for the illustration in \cref{fig:kaleidoscope}.
Finally, the authors would like to recognize the crossroads of Massachusetts Ave and Langdon St as the origin of the neutrino kaleidoscope.
N.W.K. is supported by the David and Lucile Packard Foundation and NSF IAIFI. G.P. is supported by the Lederman Fellowship at Fermilab. This work was produced by FermiForward Discovery Group, LLC under Contract No. 89243024CSC000002 with the U.S. Department of Energy, Office of Science, Office of High Energy Physics. The United States Government retains and the publisher, by accepting the work for publication, acknowledges that the United States Government retains a non-exclusive, paid-up, irrevocable, world-wide license to publish or reproduce the published form of this work, or allow others to do so, for United States Government purposes. The Department of Energy will provide public access to these results of federally sponsored research in accordance with the DOE Public Access Plan (http://energy.gov/downloads/doe-public-access-plan).

\section*{ENDMATTER}

\subsection{Active Neutrino Oscillation Reach on Anomalous Matter Potentials}

In addition to Lorentz invariance violation, $\nu_\mu-\nu_\tau$ oscillations would also be able to probe other non-standard oscillation scenarios. For example, the oscillation could be generated by a matter potential experienced by the neutrinos traveling through the earth~\cite{NeutrinoNSI1, NeutrinoNSI2, NeutrinoNSI3, NeutrinoNSIDUNE, AtmospericNeutrinoNSI}. Non-LIV oscillation scenarios would not induce a sidereal variation. However, in addition to $\nu_\mu$ disappearance, these scenarios could also be identified by anomalous $\nu_\tau$ appearance. Neutrino kaleidoscopes would need to search for $\nu_\tau$ interactions on top of large $\nu_\mu$ and $\nu_e$ backgrounds.
This may be possible at the TeV scale by searching for muon and neutron ``echoes'' in hadronic cascades~\cite{Li:2016kra}. In addition, $\mu$-decaying $\nu_\tau$ interactions would have higher inelasticity than $\nu_\mu$ interactions due to the missing energy from the neutrino in the $\tau$ decay, which could be used to identify $\nu_\tau$ interactions statistically.

\begin{figure}[t]
    \centering
    \includegraphics[width=0.49\textwidth]{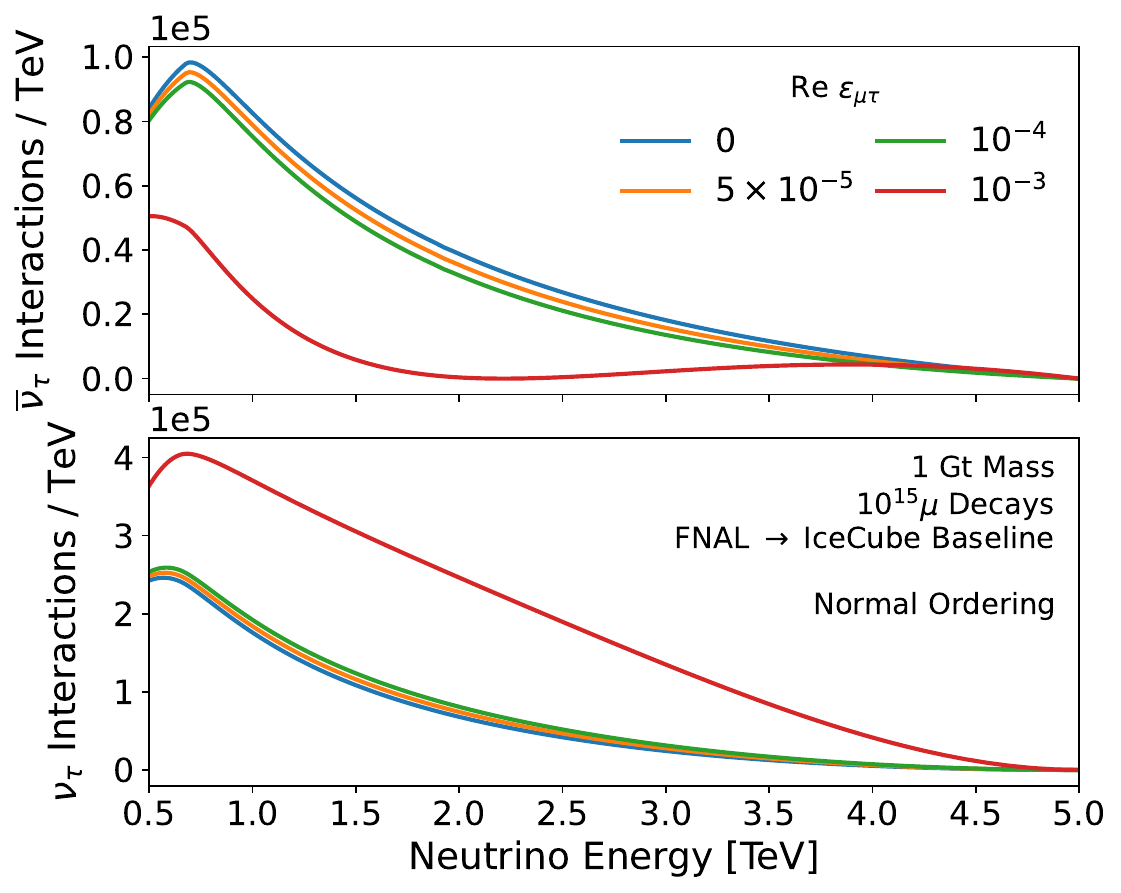}
    \caption{Interaction rate of $\overline{\nu}_\tau$ (top) and $\nu_\tau$ (bottom) for standard oscillations ($\epsilon_{\mu\tau} = 0$), as well as a few non-standard scenarios.}
    \label{fig:numutau_osc_potential}
\end{figure}

Fig.~\ref{fig:numutau_osc_potential} shows the rate of $\nu_\tau$ and $\overline{\nu}_\tau$ interactions at a range of anomalous matter potential values ($\epsilon_{\mu\tau}$, where $V_{\mu\tau} \equiv \sqrt{2}G_F N_e \epsilon_{\mu\tau}$), as well as for standard oscillations ($\epsilon_{\mu\tau} = 0$). There would be hundreds of thousands of $\nu_\tau$ interactions in the telescope to observe such an oscillation effect. However, due to the challenge of selecting a pure sample of $\nu_\tau$ events, it is difficult to estimate how well the rate of interactions in the telescope could be determined. A value of $\epsilon_{\mu\tau} = 5\times10^{-5}$ has an $\mathcal{O}($10\%$)$ effect on the $\nu_\tau$ rate, differing in direction between tau neutrinos and anti-neutrinos.
A sensitivity to this magnitude would surpass the reach on any non-standard matter potential term from existing and planned measurements~\cite{NeutrinoNSI1, NeutrinoNSI2, NeutrinoNSI3, NeutrinoNSIDUNE, AtmospericNeutrinoNSI}. The relative direction of the change in the interaction rate flips if the sign of the potential flips or, equivalently, in the case of the inverted mass ordering. An imaginary $\epsilon_{\mu\tau}$ would instead impact the $\nu_\tau$ and $\overline{\nu}_\tau$ rates in the same direction.

Neutrino non-standard matter potential terms arise naturally in radiative neutrino mass models~\cite{NuMassNSI} such as the Zee model~\cite{Zee}. Such potentials generically have a magnitude on the order of $M_W^2 / M_\chi^2$~\cite{NeutrinoNSI1}, where $M_W$ is the mass of the W boson and $M_\chi$ is the new physics scale. A sensitivity to $\epsilon_{\mu\tau} = 5\times10^{-5}$, therefore, translates into a reach on new physics generating the neutrino masses up to about \SI{10}{TeV}. 

\subsection{Rare Neutrino Interactions}

\begin{figure}
    \centering
    \includegraphics[width=0.49\textwidth]{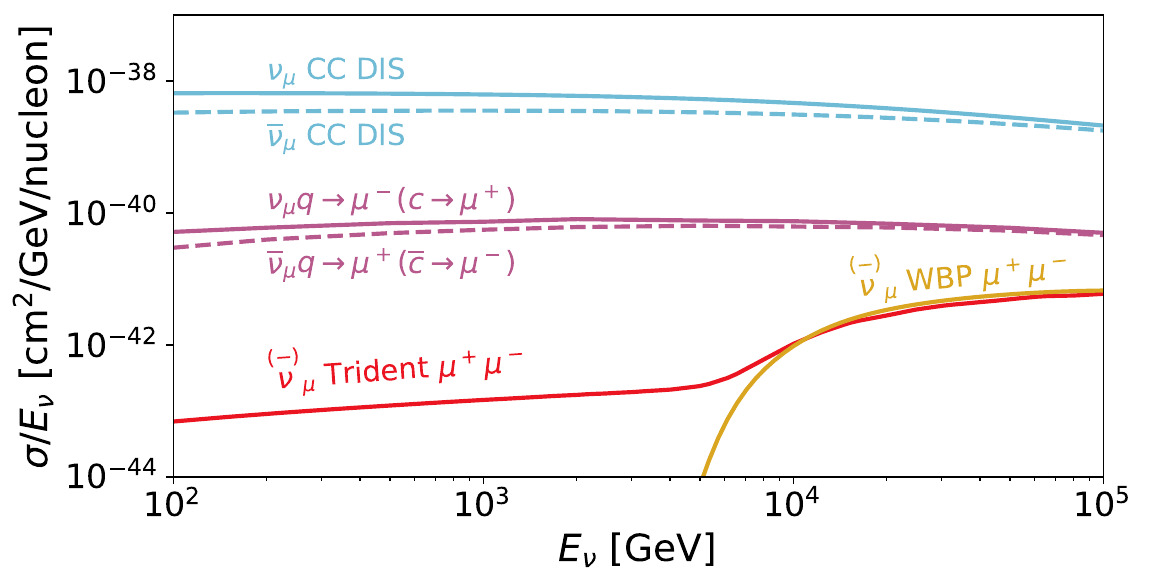}
    \caption{The total CC DIS cross section for muon neutrinos~\cite{Cooper-Sarkar:2011jtt}, and cross sections for rare neutrino interactions producing dimuons in the final state, through charm production~\cite{Weigel:2024gzh,DeLellis:2004ovi}, tridents, and WBP~\cite{Zhou:2019vxt}.}
    \label{fig:cross_sections}
\end{figure}

Neutrino kaleidoscopes can also search for rare neutrino interactions.
Specifically, neutrino production of muon pairs can lead to double track signatures in neutrino telescopes~\cite{Zhou:2021xuh}, which can in principle be distinguished from the typical track and cascade signatures.
These ``dimuon'' events can arise from charm production, W-boson production (WBP), and trident interactions.
The cross sections for these processes are shown in \cref{fig:cross_sections}.
We use the charged-current $\nu_\mu$ charm-production cross section computed in Ref.~\cite{Weigel:2024gzh} and account for the muon branching ratio of the outgoing charm hadron according to the fragmentation fractions of the outgoing charm quark~\cite{DeLellis:2004ovi}.
W-boson production and trident dimuon cross sections are taken from Refs.~\cite{Zhou:2019vxt,Zhou:2019frk}.
\Cref{tab:dimuon_rate} reports the expected rate per year from these dimuon processes in each kaleidoscope configuration.
These large dimuon event rates could be powerful probes of the strong force~\cite{Cruz-Martinez:2023sdv} and new lepton flavor symmetries~\cite{Altmannshofer:2014pba}.
We leave a detailed investigation of this possibility to future work.

\begin{table}[]
    \centering
    \begin{tabular}{|c|c|c|c|}
Process & KM3NeT & IceCube & P-ONE (IP) \\
\hline
$\nu_\mu$ CC DIS & $1.7 \times 10^{8}$ & $1.6 \times 10^{8}$ & $1.3 \times 10^{11}$ \\
$\nu_\mu q \to \mu^- (c \to \mu ^+)$ & $2.2 \times 10^{6}$ & $2.1 \times 10^{6}$ & $1.8 \times 10^{9}$ \\
${\nu}_\mu$ Trident $\mu^+ \mu^-$ & $5.2 \times 10^{3}$ & $5 \times 10^{3}$ & $4.2 \times 10^{6}$ \\
${\nu}_\mu$ WBP $\mu^+ \mu^-$ & $3.6$ & $3.6$ & $2.8 \times 10^{3}$ \\
    \end{tabular}
    \caption{The number of $\nu_\mu$ interactions per year, including rare interactions with dimuons in the final state, for each kaleidoscope configuration. Event rates are computed using the cross sections in \cref{fig:cross_sections}.} 
    \label{tab:dimuon_rate}
\end{table}

\bibliography{bib}


\clearpage
\newpage

\onecolumngrid
\appendix

\ifx \standalonesupplemental\undefined
\setcounter{page}{1}
\setcounter{figure}{0}
\setcounter{table}{0}
\setcounter{equation}{0}
\fi

\renewcommand{\thepage}{Supplemental Methods and Tables -- S\arabic{page}}
\renewcommand{\figurename}{SUPPL. FIG.}
\renewcommand{\tablename}{SUPPL. TABLE}
\renewcommand{\theequation}{A\arabic{equation}}

\crefname{figure}{Supplemental Figure}{Supplemental Figures}
\crefname{table}{Supplemental Table}{Supplemental Tables}
\Crefname{figure}{Supplemental Figure}{Supplemental Figures}
\Crefname{table}{Supplemental Table}{Supplemental Tables}


\section{More details on the Lorentz Violation Sensitivity} \label{sec:lv_app}

\subsection{Theory}

The imprint of Planck-scale Lorentz violation on the Standard Model is described by an effective field theory, the Standard-Model Extension (SME)~\cite{Colladay:1996iz,Colladay:1998fq}, in which neutrino flavor evolution is governed by the effective Hamiltonian $h_{\rm eff}$ such that
\begin{equation} \label{eq:lv}
\begin{split}
    &i\frac{d}{dx}\begin{pmatrix}
        \nu_\alpha\\
        \nu_\beta
    \end{pmatrix} =
    \begin{bmatrix}
        (h_{\rm eff})_{\alpha\alpha} & (h_{\rm eff})_{\alpha\beta} \\ 
        (h_{\rm eff})^*_{\alpha\beta} & (h_{\rm eff})_{\beta\beta}
    \end{bmatrix}
    \begin{pmatrix}
        \nu_\alpha\\
        \nu_\beta
    \end{pmatrix}, \\
     &(h_{\rm eff})_{\alpha\beta} = \frac{1}{E}[(a_L)^\mu_{\alpha \beta} p_\mu - (c_L)^{\mu \nu}_{\alpha \beta} p_\mu p_\nu] ,
\end{split}
\end{equation}
where $p_\mu$ is the neutrino four-momentum, and $(a_L)^\mu_{\alpha \beta}$ and $(c_L)^{\mu \nu}_{\alpha \beta}$ are the flavor-dependent SME coefficients associated with operators of mass dimension 3 and 4, respectively.
Here we have assumed neutrino oscillations from mass-mixing are negligible compared to those from Lorentz-violation, which is valid when $\Delta m^2 / (4E) \ll h_{\rm eff}$.
When the baseline $L$ is small compared to $h_{\rm eff}/(\hbar c)$, neutrino oscillation probabilities are given by 
\begin{equation}
P_{\nu_\alpha \to \nu_\beta} \approx |(h_{\rm eff})_{\alpha \beta}|^2 L^2 / (\hbar c)^2,
\end{equation}
where $\alpha \neq \beta$.

Two effects are important to note in \cref{eq:lv}.
First, the $a_L$ and $c_L$ coefficients have mass dimension 1 and 0, resulting in neutrino oscillations that depend on $L$ and $LE$, respectively, in contrast to the standard $L/E$ dependence of mass-generated oscillations.
Oscillations from Lorentz violation can thus dominate over mass-generated oscillations at high energies, even for Planck-suppressed SME coefficients.
Second, the non-trivial dependence on $p_\mu$ leads to frame-dependent oscillations.
As discussed in \cref{sec:lv}, the conventional choice is a Sun-centered celestial-equatorial frame, in which the Earth's rotation causes the direction of a neutrino beam to vary over the course of a sidereal day, i.e. with a frequency $\omega_{\oplus} = 2\pi/(23~{\rm h}~56~{\rm min})$.
This leads to the explicit sidereal time dependence in the effective Hamiltonian as shown in \cref{eq:lv_sidereal} and reproduced here,
\begin{equation} \label{eq:app_lv_sidereal}
\begin{split}
(h_{\rm eff})_{\alpha \beta} = 
&(\mathcal{C})_{\rm \alpha \beta}
 +  (\mathcal{A}_s)_{\rm \alpha \beta} \sin \omega_\oplus T_\oplus 
 +  (\mathcal{A}_c)_{\rm \alpha \beta} \cos \omega_\oplus T_\oplus \\
 +  &(\mathcal{B}_s)_{\rm \alpha \beta} \sin  2 \omega_\oplus T_\oplus 
 +  (\mathcal{B}_c)_{\rm \alpha \beta} \cos 2 \omega_\oplus T_\oplus,
 \end{split}
\end{equation}
where $\mathcal{C}$, $\mathcal{A}_s$, $\mathcal{A}_c$, $\mathcal{B}_s$, and $\mathcal{B}_c$ are energy-dependent linear combinations of the $a_L^\mu$ and $c_L^{\mu \nu}$ coefficients.
The specific linear combination depends on the geometry of the neutrino beam with respect to the Earth $\vec{N}$; explicit expressions can be found in Eq.~4-13 of Ref.~\cite{Kostelecky:2004hg}.
The directional vectors in the Sun-centered celestial-equatorial frame for each kaleidoscope configuration are
\begin{equation}
\begin{pmatrix}
    \hat{N}_X \\
    \hat{N}_Y \\
    \hat{N}_Z \\
\end{pmatrix}_{\rm KM3}
=
\begin{pmatrix}
    0.1836 \\
    0.5876 \\
    0.788 \\
\end{pmatrix},~~~~~
\begin{pmatrix}
    \hat{N}_X \\
    \hat{N}_Y \\
    \hat{N}_Z \\
\end{pmatrix}_{\rm IC}
=
\begin{pmatrix}
    0.0124 \\
    0.4079 \\
    -0.9129 \\ 
\end{pmatrix},~~~~~
\begin{pmatrix}
    \hat{N}_X \\
    \hat{N}_Y \\
    \hat{N}_Z \\
\end{pmatrix}_{\rm P-ONE}
=
\begin{pmatrix}
    0.1099 \\
    -0.9591  \\
    0.2609 \\
\end{pmatrix}.
\end{equation}

\subsection{Analysis}

\begin{figure}
    \centering
    \includegraphics[width=0.48\textwidth]{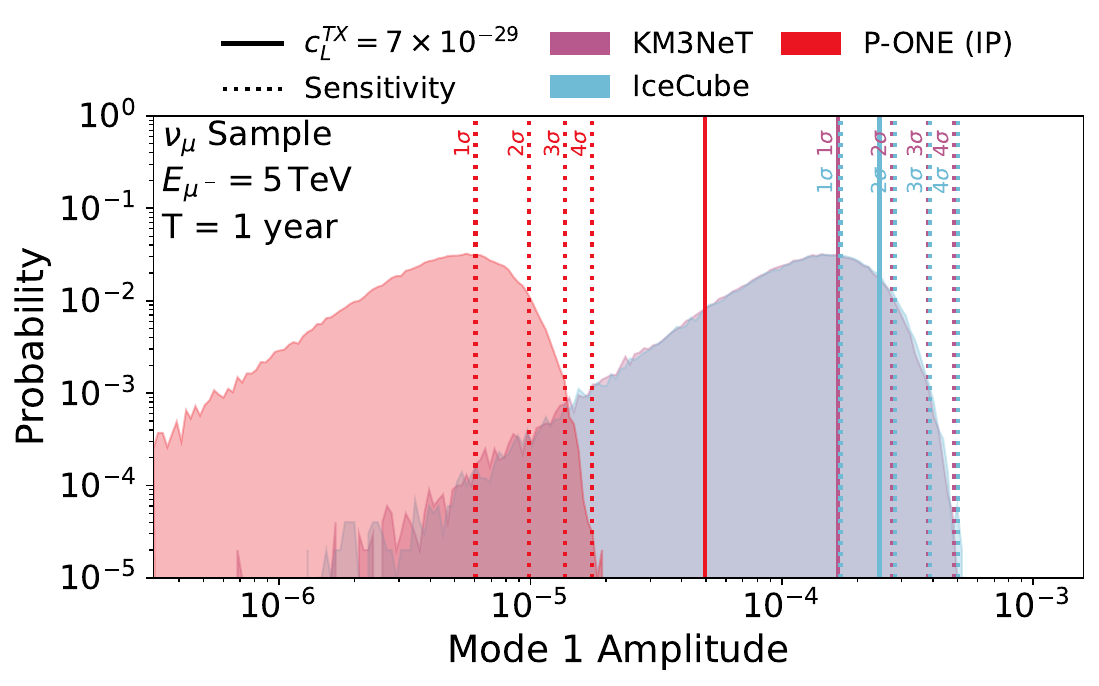}
    \includegraphics[width=0.48\textwidth]{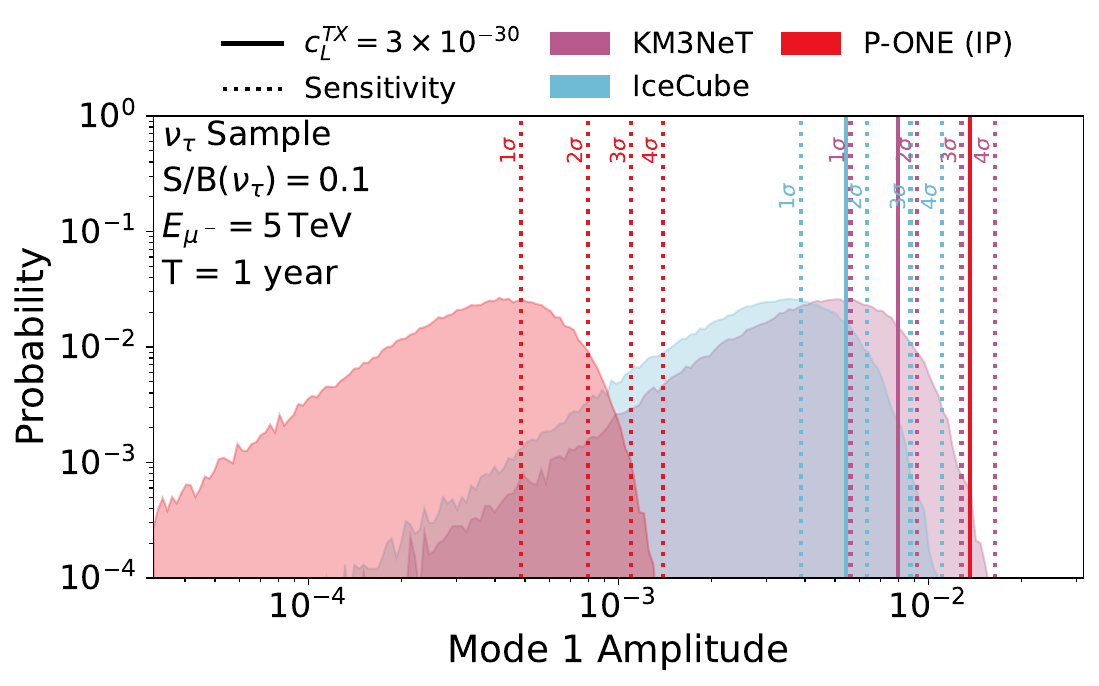}
    \caption{Expected distributions of first mode amplitudes under the null hypothesis for the $\nu_\mu$ (left) and $\nu_\tau$-enriched (right) sample in each experimental configuration, with sensitivities indicated by the dotted lines. The solid lines indicated the expected first harmonic amplitude for the indicated value of $c_L^{TX} = 1$. Here we consider one year of data at a muon energy of $5\,{\rm TeV}$.}
    \label{fig:app_lv_distribution}
\end{figure}

We begin by using \cref{eq:lv_sidereal} to determine the expected sidereal oscillation signal at each neutrino telescope as a function of the SME coefficients.
\Cref{fig:lv_example} shows the expected rate in the $\nu_\mu$ sample and the $\nu_\tau$-enriched sample in each kaleidoscope over one sidereal day for different values of $c_L^{TX}$, as a ratio to the Standard Model expectation.
Statistical uncertainties reflect predicted event rates after one year of neutrino kaleidoscope operation.
In the $\nu_\mu$ sample, the largest variations occur in IceCube due to the long baseline, while the variations are the same size in the $\nu_\tau$-enriched sample as the Standard Model prediction also depends linearly on the baseline.
In both samples, the smallest statistical uncertainties occur in P-ONE due to the event rate enhancement from the 100~m decay region.
For this specific Lorentz violation scenario, oscillations occur at the first harmonic of the sidereal frequency.
In general, SME-induced oscillations can occur as any integer harmonic of the sidereal frequency.
We follow the procedure outlined in Ref.~\cite{MINOS:2008fnv} to assess our sensitivity.
For each detector, we generate $10^5$ pseudo-experiments from the null hypothesis (no sidereal variations) and take the Fourier transform of the sidereal time distribution to determine the amplitude of each mode corresponding to the first four harmonics of the sidereal frequency.
The first harmonic distributions for the $\nu_\mu$ and $\nu_\tau$-enriched samples are shown in \cref{fig:app_lv_distribution}.
As we are looking for amplitudes in excess of the null expectation, the vertical dotted lines show the one-tailed sensitivity at the $\{1,2,3,4\}\sigma$ confidence level.
The solid vertical lines indicate the expected amplitude of the first harmonic for the indicated values of $c_L^{TX}$.
In this example, the $\nu_\mu$ sample of the P-ONE (IP) neutrino kaleidoscope would rule out $c_L^{TX} = 7 \times 10^{-29}$ at $>4\sigma$.

\begin{figure}
    \centering
    \includegraphics[width=0.48\textwidth]{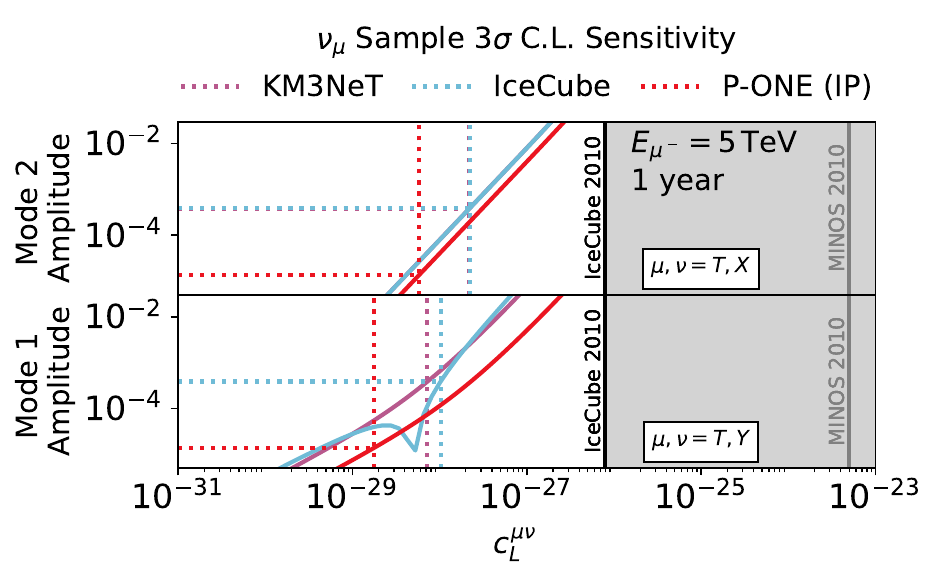}
    \includegraphics[width=0.48\textwidth]{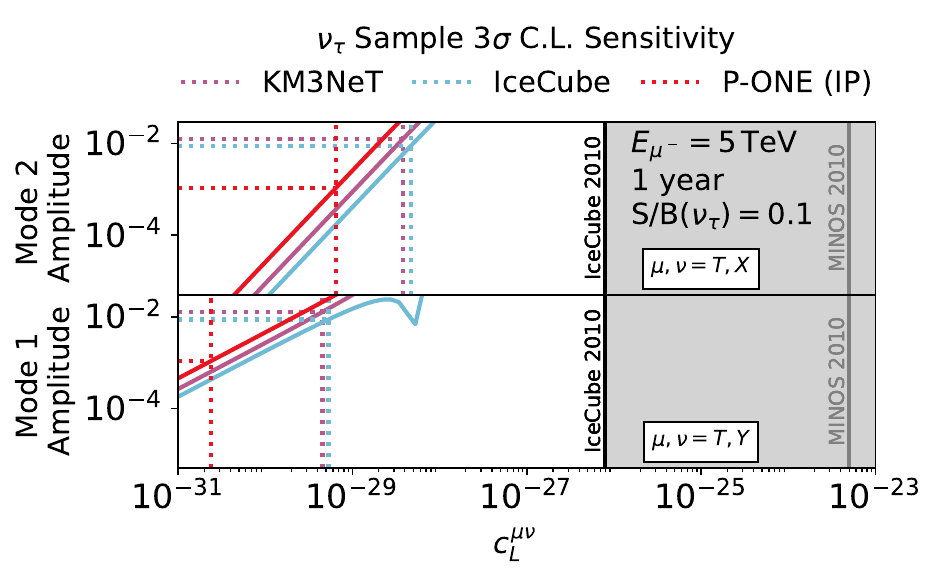}
    \caption{The expected $3\sigma$ exclusion sensitivity to $c_L^{TX}$ and $c_L^{TY}$ for the $\nu_\mu$ (left) and $\nu_\tau$-enriched (right) sample in each kaleidoscope, determined by corresponding quantile of the second and first mode amplitude distributions, respectively. Here we consider one year of data at a muon energy of $5\,{\rm TeV}$. The shaded regions correspond to existing constraints from IceCube~\cite{IceCube:2010fyu} and MINOS~\cite{MINOS:2010kat}.}
    \label{fig:app_lv_analysis}
\end{figure}

We repeat this exercise to determine the value of each SME coefficient that results in an amplitude at the $3\sigma$ sensitivity of the three neutrino kaleidoscopes.
This analysis is carried out for each of the first four harmonics of the sidereal frequency, and the strongest constraint over the four modes is taken as the $3\sigma$ sensitivity of a given kaleidoscope to a given SME coefficient.
\Cref{fig:app_lv_analysis} depicts this analysis for $c_L^{TX}$ and $c_L^{TY}$ in both the $\nu_\mu$ and $\nu_\tau$-enriched samples.
The sensitivity of the $\nu_\mu$ sample for all relevant $a_L^\mu$ and $c_L^{\mu \nu}$ coefficients is shown in \cref{tab:liv_constraints}.
As discussed in \cref{sec:lv}, neutrino kaleidoscopes can improve upon existing limits on $\mu$-$\tau$ SME coefficients by multiple orders of magnitude.

\begin{figure}
    \centering
    \includegraphics[width=0.5\textwidth]{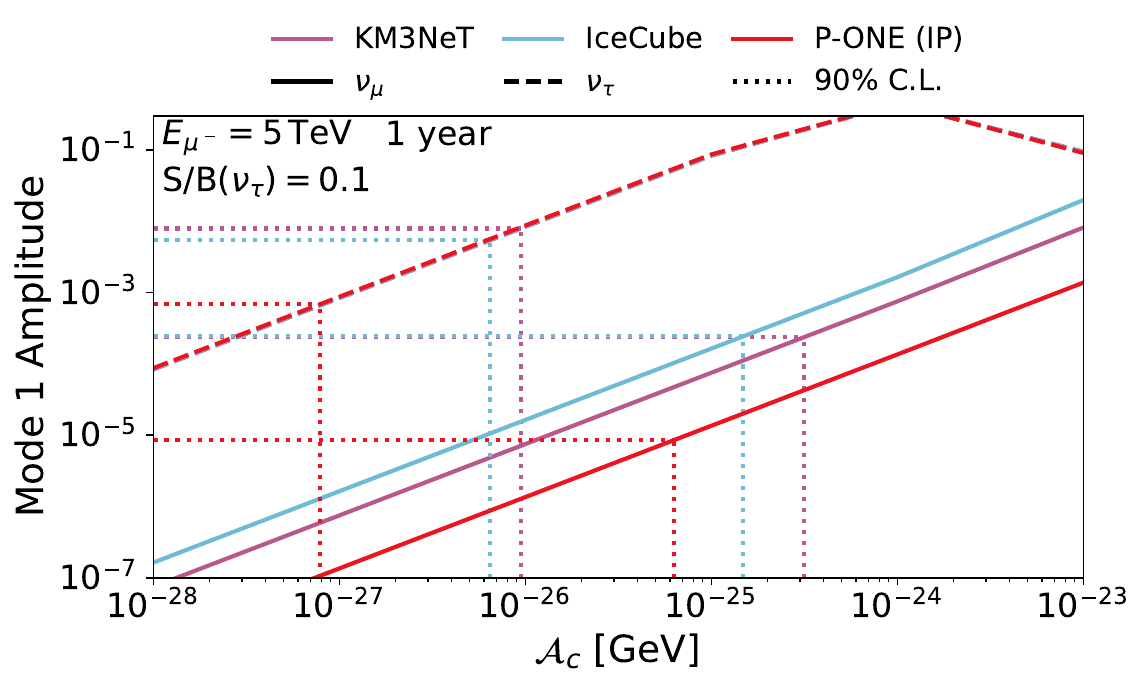}

    \caption{\textbf{Top:} The sensitivity of each kaleidoscope to $\mathcal{A}_c$ at dimension 3 for ten years of data with a muon energy of $5\,{\rm TeV}$. Separate results for a $\nu_\mu$ and $\nu_\tau$-enriched sample are shown.}
    \label{fig:app_lv_Ac_sensitivity}
\end{figure}

We now touch on analysis of higher-dimensional SME operators.
As indicated by \cref{eq:lv_higher_order}, in order to compute the sensitivity of each kaleidoscope to $A_c^d$ from operators of mass-dimension $d$, we must first compute the sensitivity to $A_c^3$ and scale by the average neutrino energy $|\mathbf{p}| \approx E_\mu/2$.
According to \cref{eq:app_lv_sidereal}, $A_c^3$ represents the amplitude of first-harmonic sidereal variations.
\Cref{fig:app_lv_Ac_sensitivity} shows the induced first harmonic amplitude as a function of $A_c^3$ for the $\nu_\mu$ and $\nu_\tau$-enriched sample in each kaleidoscope.
The dotted lines in \cref{fig:app_lv_Ac_sensitivity} indicate the $90\%$~C.L. sensitivity to $A_c^3$ computed using the psuedoexperiments described above.
\Cref{tab:app_lv_higher_order} shows the sensitivity of the nominal, tau-enriched, and high-energy kaleidoscope configurations introduced in \cref{sec:lv} to $A_c^d$, for $d \in \{3,4,5,6\}$.
The $d=6$ sensitivities in particular suggest sensntivity near the double-Planck suppresesed regime.

\begin{table}[]
    \centering
    \begin{tabular}{|c|c|c|c|c|}
Coefficient & KM3NeT & IceCube & P-ONE (IP) & Current Limit \\
\hline
$a_L^X / 10^{-25}\,{\rm GeV}$ & 8.2 & 5.8 & 1.0 & $1.7 \times 10^{2}$ (\cite{IceCube:2010fyu}) \\
$a_L^Y / 10^{-25}\,{\rm GeV}$ & 8.2 & 5.8 & 1.0 & $1.7 \times 10^{2}$ (\cite{IceCube:2010fyu}) \\
$c_L^{TX} / 10^{-28}$ & 1.6 & 1.1 & 0.2 & $79$ (\cite{IceCube:2010fyu}) \\
$c_L^{TY} / 10^{-28}$ & 0.7 & 1.0 & 0.2 & $79$ (\cite{IceCube:2010fyu}) \\
$c_L^{XX} / 10^{-28}$ & 4.0 & 2.4 & 0.7 & $2.5 \times 10^{5}$ (\cite{MINOS:2010kat}) \\
$c_L^{XY} / 10^{-28}$ & 5.2 & 5.4 & 0.4 & $1.2 \times 10^{5}$ (\cite{MINOS:2010kat}) \\
$c_L^{XZ} / 10^{-28}$ & 2.0 & 1.2 & 0.7 & $7 \times 10^{4}$ (\cite{MINOS:2010kat}) \\
$c_L^{YY} / 10^{-28}$ & 4.2 & 2.4 & 0.6 & $2.4 \times 10^{5}$ (\cite{MINOS:2010kat}) \\
$c_L^{YZ} / 10^{-28}$ & 2.0 & 1.2 & 0.7 & $7 \times 10^{4}$ (\cite{MINOS:2010kat}) \\
    \end{tabular}
    \caption{The sensitivity of the $\nu_\mu$ sample in each kaleidoscope configuration to all relevant $\mu$-$\tau$ $a_L^\mu$ and $c_L^{\mu \nu}$ SME coefficients at the $3\sigma$ C.L., considering one year of data at a muon energy of $5\,{\rm TeV}$. 
    The current limits on each coefficient are also shown, which come from sidereal variation searches at either IceCube~\cite{IceCube:2010fyu} or MINOS~\cite{MINOS:2010kat}.}
    \label{tab:liv_constraints_full}
\end{table}

\begin{table}[]
    \centering
    \begin{tabular}{|c|c|c|c|c|}
\hline
Dimension & $\mathcal{A}_c^d$ Units & KM3NeT & IceCube & P-ONE \\
\hline
\multicolumn{5}{|c|}{$\nu_\mu$ Sample, $E_\mu=5~{\rm TeV}$, 1 year} \\
\hline
$d=3$ & ${\rm GeV}^{1}$ & $3.1 \times 10^{-25}$ & $1.5 \times 10^{-25}$ & $6.3 \times 10^{-26}$ \\
$d=4$ & ${\rm GeV}^{0}$ & $1.2 \times 10^{-28}$ & $5.7 \times 10^{-29}$ & $2.3 \times 10^{-29}$ \\
$d=5$ & ${\rm GeV}^{-1}$ & $4.7 \times 10^{-32}$ & $2.2 \times 10^{-32}$ & $8.7 \times 10^{-33}$ \\
$d=6$ & ${\rm GeV}^{-2}$ & $1.8 \times 10^{-35}$ & $8.3 \times 10^{-36}$ & $3.2 \times 10^{-36}$ \\
\hline
\multicolumn{5}{|c|}{$\nu_\tau$ Sample, S/B$(\nu_\tau)=0.1$, $E_\mu=5~{\rm TeV}$, 10 years} \\
\hline
$d=3$ & ${\rm GeV}^{1}$ & $3 \times 10^{-27}$ & $2.1 \times 10^{-27}$ & $2.4 \times 10^{-28}$ \\
$d=4$ & ${\rm GeV}^{0}$ & $1.2 \times 10^{-30}$ & $7.8 \times 10^{-31}$ & $9.4 \times 10^{-32}$ \\
$d=5$ & ${\rm GeV}^{-1}$ & $4.5 \times 10^{-34}$ & $2.9 \times 10^{-34}$ & $3.7 \times 10^{-35}$ \\
$d=6$ & ${\rm GeV}^{-2}$ & $1.7 \times 10^{-37}$ & $1.1 \times 10^{-37}$ & $1.4 \times 10^{-38}$ \\
\hline
\multicolumn{5}{|c|}{$\nu_\mu$ Sample, $E_\mu=50~{\rm TeV}$, 10 years} \\
\hline
$d=3$ & ${\rm GeV}^{1}$ & $1.2 \times 10^{-24}$ & $5.4 \times 10^{-25}$ & $2.5 \times 10^{-25}$ \\
$d=4$ & ${\rm GeV}^{0}$ & $4.8 \times 10^{-29}$ & $2.2 \times 10^{-29}$ & $9.2 \times 10^{-30}$ \\
$d=5$ & ${\rm GeV}^{-1}$ & $1.9 \times 10^{-33}$ & $8.8 \times 10^{-34}$ & $3.4 \times 10^{-34}$ \\
$d=6$ & ${\rm GeV}^{-2}$ & $7.3 \times 10^{-38}$ & $3.6 \times 10^{-38}$ & $1.3 \times 10^{-38}$ \\
\hline
    \end{tabular}
    \caption{The sensitivity of each kaleidoscope configuration to $\mathcal{A}_c^d$ at the 90\% C.L., considering a $\nu_\mu$ sample with one year of data at a muon energy of $5\,{\rm TeV}$.} 
    \label{tab:app_lv_higher_order}
\end{table}

\end{document}